\documentclass[twocolumn,aps,preprintnumbers,prd,superscriptaddress,10pt]{revtex4-1}
\usepackage{amsmath,amssymb,amsfonts,mathtools}
\usepackage{epsfig}
\usepackage{graphicx}
\usepackage{slashed}
\usepackage{color}
\usepackage[dvipsnames]{xcolor}
\usepackage[normalem]{ulem}

\begin{document}

\title{TMD Soft Function from Large-Momentum Effective Theory}

\author{Xiangdong Ji}
\affiliation{Tsung-Dao Lee Institute, Shanghai Jiao Tong University, Shanghai 200240, China}
\affiliation{Department of Physics, University of Maryland, College Park, MD 20742, USA}

\author{Yizhuang Liu}
\email{yizhuang.liu@sjtu.edu.cn, corresponding author}
\affiliation{Tsung-Dao Lee Institute, Shanghai Jiao Tong University, Shanghai 200240, China}

\author{Yu-Sheng Liu}
\affiliation{Tsung-Dao Lee Institute, Shanghai Jiao Tong University, Shanghai 200240, China}

\date{\today}

\begin{abstract}
We study Euclidean formulations of the transverse-momentum-dependent (TMD) soft function, which is a cross section for soft gluon
radiations involving color charges moving in two conjugate lightcone directions in quantum chromodynamics.
We show it is related to a special form factor of a pair of color sources traveling with nearly-lightlike velocities, which can be matched
to TMD physical observables in semi-inclusive deep-inelastic scattering and Drell-Yan process
in the framework of large momentum effective theory.
It can also be extracted by combining a large-momentum form factor of light meson and its leading TMD wave function. These
formulations are useful for initiating nonperturbative calculations of this useful quantity.
\end{abstract}

\maketitle
{\it Introduction.}---Radiation of soft gluons from fast-moving charged particles is a ubiquitous phenomenon of high energy processes in quantum chromodynamics (QCD).
Soft radiation usually cancels in total cross section such as inclusive deep inelastic scattering (DIS). However, for certain processes where a small transverse momentum is measured, such cancellation can be incomplete and result in measurable consequences.
The soft radiation, from arbitrary numbers of gluons at small momentum below $\Lambda_{\rm QCD}$, usually involves formidable nonperturbative physics.
Fortunately, in some cases soft radiations factorizes from other energy scales of the process which leads to surprising simplification. Intuitively, soft gluons have no impact on the velocity of the fast-moving color charged partons, and the propagators of partons eikonalize to straight gauge links along their moving trajectory.
In such cases, universal and process-independent matrix elements~\cite{Collins:1981uk,Kidonakis:1998bk,Collins:2004nx,Bauer:2008jx}, called soft functions, emerges to capture soft gluon effects.

The transverse-momentum-dependent (TMD) soft function (simply called soft function thereafter) interested in this paper appears in factorization theorems for the Drell-Yan (DY) process~\cite{Collins:1984kg,Collins:1988ig} and semi-inclusive DIS (SIDIS)~\cite{Ji:2004wu,Ji:2004xq}.
Here the soft function $S(b_\perp,\mu,Y)$ is defined as vacuum expectation value of a Wilson loop composed of
lightlike gauge links, and usually depends on three variables: the rapidity regulator $Y$, the transverse separation $b_\perp$ (conjugate to transverse momentum), and the renormalization scale $\mu$ associated to the cusps of gauge links.
Due to the gluon radiation collinear to the gauge links,
$S$ contains the well-known rapidity divergences, also called lightcone divergences or singularities.
For such singularities, a rapidity regulator is needed and its renormalization results in a rapidity scale
whose evolution is governed by the so-called Collins-Soper kernel~\cite{Collins:1981uk}.
We broadly consider two types of rapidity regulators:
(1) On-lightcone ($\delta$): The Wilson line runs strictly along the lightlike directions such as $\eta$~\cite{Chiu:2012ir}, $\delta$~\cite{Echevarria:2015byo}, exponential~\cite{Li:2016ctv} and analytical~\cite{Becher:2010tm} regularizations; (2) Off-lightcone ($Y$): The Wilson line is either
space- or time-like, asymptotically approaching the lightcone.
In the on-lightcone scheme, the soft function has been studied extensively in perturbation theory on the structure of rapidity divergences~\cite{Vladimirov:2017ksc} as well as by calculations up to three loops~\cite{Echevarria:2015byo,Luebbert:2016itl,Li:2016axz,Li:2016ctv,Luo:2019hmp}.

Despite its important role in the factorization theorems of TMD physical observables, a first-principle calculation of the soft function in the nonperturbative region (large $b_\perp$) remains an open question, due to its origin from the square of an S-matrix.
Recently, there have been attempts to formulate the soft function on lattice~\cite{Ji:2014hxa,Ji:2018hvs,Ebert:2019okf}, however,
the suggested Euclidean quantities so far cannot be used in actual TMD factorization.
In this paper, we show that the soft function in the off-lightcone scheme
can be obtained from a form factor of a pair of moving color sources, as well as by combining
a special form factor of a light meson with opposite large momenta and its TMD wave function (TMDWF). The result
can be used for nonperturbative calculations of the soft function and its
matching to TMD physical observables in DIS and DY using large-momentum effective theory (LaMET)~\cite{Ji:2013dva,Ji:2014gla}.

We introduce a soft function equivalent to the one defined by Collins~\cite{Collins:2011zzd,Collins:2011ca} and show that it is equal to the form factor of a boosted heavy-quark pair
\begin{align}\label{eq:S_HQ_mu}
S(b_\perp,Y,Y')= {}_{v'}\langle\overline QQ,\vec{b}_\perp|J(\vec{b}_\perp,v',v)|\overline QQ,{\vec b}_\perp\rangle_v
\end{align}
where $v^\mu=\gamma(1,\beta,\vec 0_\perp)$ and $v'^\mu=\gamma'(1,-\beta',\vec 0_\perp)$ are two opposite nearly-lightlike velocities in $(t,z,\vec b_\perp)$ coordinates;
$\gamma$ and $\gamma'$ are boost factors for the four velocities $v$ and $v'$, and $\beta$ and $\beta'$ are corresponding speeds;
the rapidity $Y$ and the speed $\beta$ are related through $\beta=\tanh Y$;
the renormalization scale of the soft function $S$ is suppressed for simplicity;
$|\overline QQ,\vec{b}_\perp\rangle_v$ is a heavy quark-anti-quark ground state with fixed transverse separation ${\vec b}_\perp$ and boosted to velocity $v$;
$\overline Q$ and $Q$ are color sources in heavy quark effective theory (HQET);
$J$ is the time-independent transition current defined in Eq.~(\ref{eq:J}), the cusp anomalous dimension of which is the same as for the Wilson lines \cite{Korchemsky:1987wg,Korchemsky:1991zp,Ji:1991az,Grozin:2015kna}.
The soft function actually depends on relative rapidity, $S(b_\perp,Y,Y')=S(b_\perp,Y+Y')$, due to Lorentz invariance.
We choose to keep the variable $Y$ and $Y'$ in $S$ separately to distinguish the shape of the soft function on different sides.
Equation~(\ref{eq:S_HQ_mu}) paves the way for direct lattice calculations using HQET~\cite{Aglietti:1993hf,Hashimoto:1995in,Horgan:2009ti}
or other Euclidean field-theory methods. \\

\begin{figure}
\includegraphics[width=0.85\columnwidth]{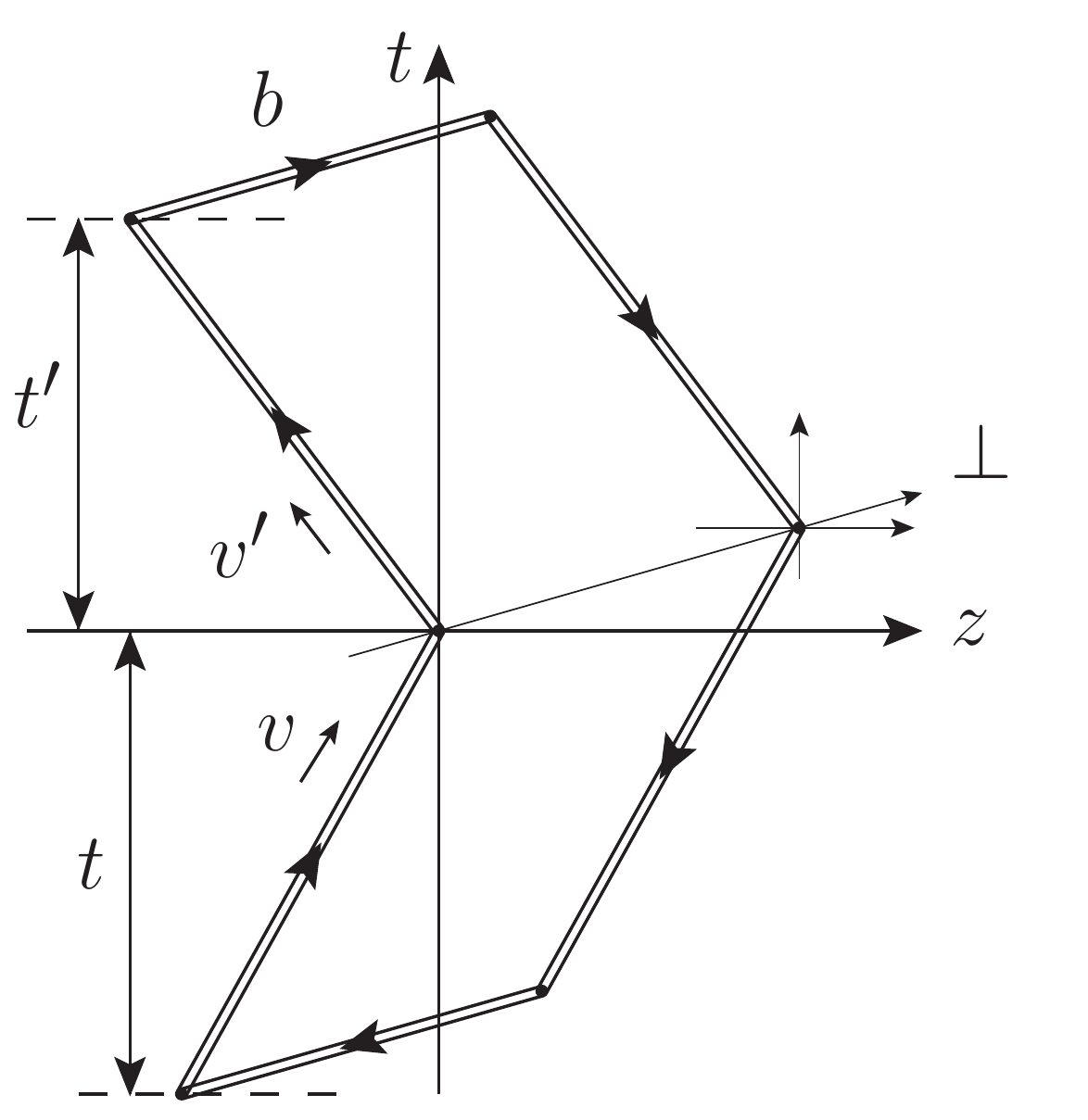}
\caption{\label{fig:S_t} The Wilson loop defined in Eq.~(\ref{eq:W_t}): The double line represents gauge links in Minkowski space.}
\end{figure}

{\it Off-lightcone soft function.}---We define a scattering amplitude of Wilson loop as showing in Fig.~\ref{fig:S_t}:
\begin{align}\label{eq:W_t}
&W(t,t',b_\perp,Y,Y')\nonumber \\
&=\frac{1}{N_c}{\rm Tr\,}\langle\Omega|{\cal T} \left[{\cal W}^{\dagger}_{v'}(\vec{b}_\perp,t'){\cal W}_{v}(\vec{b}_\perp,t)\right]|\Omega\rangle
\end{align}
where $|\Omega\rangle$ is the QCD vacuum state;
$N_c$ is number of colors and ${\rm Tr}$ is the color-trace;
timelike four-vectors $v^\mu=\gamma(1,\beta,\vec 0_\perp)$ and $v'^\mu=\gamma'(1,-\beta',\vec 0_\perp)$ approach lightcone as $\beta$ and $\beta' \to 1$;
$\cal T$ is a time-order operator;
${\cal W}_{v}(\vec{b}_\perp,t)$ is a staple shaped gauge-link along $v$ direction;
$t$ and $t'$ are the lengths of the $t$-components of the staples;
the staple shaped gauge-link for a generic four vector $\xi$ is defined as ${\cal W}_v(\xi,t)=W_v(0,t)W_\perp W^{\dagger}_v(\xi,t)$ where $W_v(\xi,t)={\cal P}{\rm exp}\left[-ig\int^0_{-t/\gamma} ds v\cdot A(\xi+sv)\right]$ is a gauge-link along $v$ direction, and $W_\perp$ is a transverse gauge-link at time $t$ to maintain gauge invariance.
The single time-order prescription for $S$ allows physical interpretation as a chronological process.

The off-lightcone soft function in Eq.~(\ref{eq:W_t}) contains pinch-pole singularities associated to time evolution of initial and final states at large $t$ and $t'$.
However, such singularities and self-interaction of the staples will be cancelled in the factorization formula for cross sections.
Thus, it is consistent to subtract them from the beginning in Eq.~(\ref{eq:W_t}) with rectangular Wilson loop~\cite{Collins:2008ht,Ji:2018hvs}.
This leads to the definition of the soft function:
\begin{align}\label{eq:S_t}
S(b_\perp,Y,Y')=\lim_{\substack{t\to\infty\\t'\to\infty}}\frac{W(t,t',b_\perp,Y,Y')}{\sqrt{Z(2t,b_\perp,Y)Z(2t',b_\perp,Y')}}
\end{align}
where $Z(2t,b_\perp,Y)$ is the vacuum expectation of rectangular Wilson loop which is similar to $W$ by setting $v'=v$ and $t'=t$, i.e. $Z(2t,b_\perp,Y)=W(t,t,b_\perp,Y,-Y)$.
The factor $Z$ has a clear physical interpretation: It can be viewed as the wave function renormalization for incoming or outgoing color sources.
After the subtraction through $Z$, the only remaining divergences for $S(b_\perp,Y,Y')$ are cusp divergences with hyperbolic angle $Y+Y'$.

For large rapidities, the soft function can be expressed as~\cite{Collins:1981uk,Vladimirov:2017ksc}
\begin{align}\label{eq:S_diff_scheme}
S(b_\perp,Y,Y')=e^{(Y+Y')K(b_\perp)+{\cal D}(b_\perp)+{\cal O}(\exp[-(Y+Y')])}
\end{align}
where $Y$ and $Y'$ are off-lightcone regulators; similarly in the on-lightcone scheme, $\delta$ and $\delta'$ are used as corresponding regulators.
The renormalization scales of $S$, $K$, and $\cal D$ are suppressed for simplicity.
$K(b_\perp)$ is the nonperturbative part of the Collins-Soper kernel,
which has been shown that it can be obtained through momentum evolution in quasi-TMDPDF~\cite{Ebert:2018gzl}.
$K$ is generally believed to be rapidity scheme-independent~\cite{Vladimirov:2017ksc},
however, $\cal D$ is not.
The relationship between $\cal D$'s in on- and off-lightcone schemes will be discussed after Eq.~(\ref{eq:SSS}).
Our main result is about $K$ and $\cal D$ in the off-lightcone scheme, which can be used
to obtain (or match to) scheme-independent lightcone TMD quantities in the factorization
of physical observables, such as TMDPDFs and TMDWFs.\\

{\it Soft function as form factor of color source pair.}---In HQET, the propagator of a color source is equivalent to a gauge link along its moving direction.
Thus $W(t,t',b_\perp,Y,Y')$ can be expressed by fields in HQET with the Lagrangian
\begin{align}
{\cal L_{\rm HQET}}=\psi_v^{\dagger}(x)(iv \cdot D)\psi_v(x)+\eta_v^{\dagger}(x)(iv \cdot D)\eta_v(x)
\end{align}
where $\psi_v$ and $\eta_v$ are quark and anti-quark in the fundamental and anti-fundamental representations, respectively; $v^\mu=\gamma(1,\beta,\vec 0_\perp)$ is the four velocity; $D$ is the covariant derivative.
Note that quarks in HQET can be viewed as color sources.
If the gluon soft function is considered, the heavy quarks should be in adjoint representation.

In HQET, a color-singlet heavy-quark pair separated by $\vec b$ generates a heavy quark potential $V(\vec b)$ in the ground state, and the spectrum includes a gapped continuum above it.
The state can also have a residual momentum $\delta\vec P$, which is arbitrary due to reparameterization invariance~\cite{Luke:1992cs,Manohar:2000dt}, and for simplicity we always consider $\delta\vec P=0$.
When the sources move with a velocity $v$, the ground state can be labeled by
$|\overline Q Q,\vec{b},\delta \vec P\rangle_v$, where $\delta\vec P=\vec{P}_{\rm total}-2m_Q\gamma\vec\beta$.
The residual energy of the state is $E=\gamma^{-1}V(\vec b)+\vec{\beta}\cdot \delta\vec P$.

Consider a process with incoming and outgoing states being heavy-quark pairs separated by ${\vec b}_\perp$ and at velocity $v$ and $v'$, respectively.
Such a state is created by the interpolating fields
\begin{align}
{\cal O}_v(t,{\vec b}_\perp)=\int d^3\vec r \, \psi_v^{\dagger}(t,\vec r\,) {\cal U}(\vec r,\vec r\,',t) \eta_v^{\dagger}(t,\vec r\,')
\end{align}
where $\vec r\,'=\vec r+\vec b_\perp$;
${\cal U}(\vec r,\vec r\,',t)$ is a gauge link connecting $\vec r\,'$ to $\vec r$ at time $t$.
The heavy-quark pair created by ${\cal O}_v$ is forced to be at relative separation $\vec{b}_\perp$ and to have vanishing residual momentum $\delta \vec P=0$.
Between the incoming and outgoing states, a product of two local equal-time operators
\begin{align}\label{eq:J}
J(\vec{b}_\perp,v,v')= \eta^\dagger_{v'}(\vec{b}_\perp)\eta_v(\vec{b}_\perp)\psi_{v'}^\dagger(0)\psi_v(0)
\end{align}
is inserted at $t=0$.
Then $W$ can be expressed in terms of HQET propagators, which equal to gauge links.
One contracts heavy-quark fields in the following expression, up to an overall volume factor, we obtain,
\begin{align}\label{eq:W_HQ}
&W(t,t',b_\perp,Y,Y') \nonumber\\
&=\frac{1}{N_c} \langle\Omega|{\cal O}^{\dagger}_{v'}(t',{\vec b}_\perp) J(\vec{b}_\perp,v,v') {\cal O}_v(-t,\vec{b}_\perp)|\Omega\rangle \nonumber\\
&\xrightarrow[]{\substack{t\to\infty\\t'\to\infty}} \frac{1}{N_c}\Phi^\dagger(\vec{b}_\perp) S(b_\perp,Y,Y') \Phi(\vec{b}_\perp)e^{-iE't'-iEt}
\end{align}
where
\begin{align}
\Phi(\vec{b}_\perp)&=\lim_{t\to\infty}{}_v\langle \overline QQ,\vec{b}_\perp|{\cal O}_v(t,\vec{b}_\perp)|\Omega\rangle\, ,\\
S(b_\perp,Y,Y')&={}_{v'}\langle\overline QQ,\vec{b}_\perp|J(\vec{b}_\perp,v,v')|\overline QQ,\vec{b}_\perp\rangle_v\, ,\label{eq:S_HQ}
\end{align}
and $\beta=\tanh Y$.
In the last line of Eq.~(\ref{eq:W_HQ}), we insert a complete set of heavy-quark pair states before and after $J$.
At large time, the contribution from the continuum spectrum is damped out due to the Riemann-Lebesgue lemma, while the contribution from $|\overline QQ,\vec{b}_\perp,\delta \vec{P}=0\rangle_v$ with residual energy $E=\gamma^{-1}V(\vec b_\perp)$ survives.
As a result we obtain Eqs.~(\ref{eq:W_HQ}) to (\ref{eq:S_HQ}).
We have omitted the state label $\delta \vec{P}=0$ for simplicity .
Alternatively, we can also give $t$ and $t'$ a small negative imaginary part, which is consistent with the time order, to damp out all states except $|\overline QQ,\vec{b}_\perp\rangle_v$ at large $t$.
Note that $\Phi(\vec{b}_\perp)$ is independent of $Y$ because it is boost invariant.

Similarly, $Z$ can also be formulated by HQET
\begin{align}\label{eq:Z_HQ}
Z(2t,b_\perp,Y)&=\frac{1}{N_c}\langle\Omega|{\cal O}^\dagger_v(t,{\vec b}_\perp){\cal O}_v(-t,\vec{b}_\perp)|\Omega\rangle \nonumber\\
&\xrightarrow[]{\substack{t\to\infty}}\frac{1}{N_c}\Phi^\dagger({\vec b}_\perp )\Phi({\vec b}_\perp)e^{-2iEt}
\end{align}
whose $t$-component has the length $2t$.
The $Y$ dependence of $Z$ is implicit in $E$.
Combining Eqs.~(\ref{eq:W_HQ}) and (\ref{eq:Z_HQ}), we obtain $S$ defined in Eq.~(\ref{eq:S_t}). We emphasize that Eq.~(\ref{eq:S_t}) can be seen as the Lehmann-Symanzik-Zimmermann reduction formula, in which we amputate the external heavy-quark pair states $|\overline QQ,{\vec b}_\perp\rangle_v$.

Being an equal-time observable, $S(b_\perp,Y,Y')$ can be straightforwardly realized in Euclidean time:
\begin{align}\label{eq:S_lattice}
S(b_\perp,Y,Y')=\lim_{\substack{T\to\infty\\T'\to\infty}}\frac{W_E(T,T',b_\perp,Y,Y')}{\sqrt{Z_E(2T,b_\perp,Y)Z_E(2T',b_\perp,Y')}}
\end{align}
where the subscript $E$ indicates the quantity is defined in Euclidean time, with corresponding
variables $T$ and $T'$.
The relevant matrix elements are now calculated by a lattice version of HQET with the Lagrangian~\cite{Aglietti:1993hf,Hashimoto:1995in,Horgan:2009ti}
\begin{align}
{\cal L}_{{\rm HQET},E}=\psi_v^\dagger(x) (i\tilde v\cdot D_E)\psi_v(x)+\eta_v^\dagger(x) (i\tilde v\cdot D_E)\eta_v(x)
\end{align}
where the subscript $E$ denotes the Euclidean space, $i\tilde v\cdot D_E=D^\tau-i\beta D^z$ with $\tilde v^\mu=\gamma(-i,-\beta,\vec 0_\perp)$. We have explicitly verified Eq. (\ref{eq:S_lattice}) to the one-loop order.

The soft function cannot be calculated on lattice by simply replacing the Minkowski gauge links in Eq.~(\ref{eq:W_t}) by a finite number of Euclidean gauge links. Through HQET, we find a time-independent formulation of the soft function, and open up the possibility of direct lattice calculations.\\

{\it Soft function from large-momentum light-meson form factor.}---Consider a similar large-momentum form factor as in Eq.~(\ref{eq:S_HQ}) for pseudoscalar light-meson states with constituents $\overline\psi\eta$,
\begin{align}
F(b_\perp,P\!\cdot\!P')=\langle P'|\overline\eta(\vec b_\perp)\Gamma'\eta(\vec b_\perp) \overline\psi(0)\Gamma\psi(0)|P\rangle
\end{align}
where $\psi$ and $\eta$ are light quark fields of different flavors;
$P^\mu=(P^t,P^z,\vec 0_\perp)$ and $P'^\mu=(P^t,-P^z,\vec 0_\perp)$ are two large momenta which approach to two lightlike directions in the limit $P^z\to\infty$;
$\Gamma$ and $\Gamma'$ are Dirac gamma matrices.
At large $P^z$, similar to the situation for the DY process, $F$ can be factorized in lightcone
quantities~\cite{future}
\begin{align}\label{eq:F_factorization}
F(b_\perp,P\!\cdot\!P')&=\int dx dx' H_F(x,x',P\!\cdot\! P')\\
&\times\frac{\phi(x',b_\perp,P',\overline \delta')}{S_2(b_\perp,\overline \delta',\delta')} \frac{\phi^\dagger(x,b_\perp,P,\overline \delta)}{S_2(b_\perp,\delta,\overline \delta)} S_2(b_\perp,\delta,\delta')\nonumber
\end{align}
where $H_F$ is a perturbative hard kernel.
The hard scale in $H_F$ is of the form $P\!\cdot\! P'$, which can be understood since this is the only Lorentz invariant combination of $P$ and $P'$.
$\delta$, $\delta'$, $\overline \delta$, and $\overline \delta'$ are on-lightcone rapidity regulators for gauge links along $P$, $P'$, $P'$, and $P$ directions, respectively.
The rapidity regulators are understood to take the lightcone limit.
The subscript ``2'' of the soft function denotes that both rapidity regulators are on-lightcone.
All gauge links are future-pointing and all the rapidity regulators cancel with each other to maintain manifest scheme independence.
$\phi$ is the lightcone TMDWF with on-lightcone regulator $\delta$,
\begin{align}
&\phi(x,b_\perp,P,\delta)\nonumber\\
&=\int\frac{d\lambda}{4\pi}e^{-ix\lambda }\langle P|\overline\psi(\lambda n+\vec{b}_\perp)\gamma_5\gamma^+\,{\cal W}_n\psi(0)|\Omega\rangle
\end{align}
where $n^\mu=(1,-1,\vec 0_\perp)/(P^z+P^t)$ in $(t,z,\vec b_\perp)$ coordinates;
${\cal W}_n=W_n^{\dagger}(\lambda n+\vec{b}_\perp)W_\perp W_n(0)$ is a staple shaped gauge link along $n$ direction similar to those defined in Eq.~(\ref{eq:W_t}), where $W_n(\xi)={\cal P}{\rm exp} \left[-ig\int_{0}^{\infty} n\cdot A(\xi+sn)\right] $.
The same lightcone TMDWF also appears in factorization for electromagnetic pion form factor in Ref.~\cite{Li:1992nu}.

To extract soft functions from the lattice calculable form factor in Eq.~~(\ref{eq:F_factorization}),
we need to know the lightcone TMDWF as well.
Therefore, we construct a lattice calculable quasi-TMDWF~\cite{Ji:2013dva},
\begin{align}\label{eq:quasi-TMDWF}
&\widetilde \phi(x,b_\perp,P)\\
&=\lim_{L\to\infty}\int\frac{d\lambda}{4\pi}e^{ix \lambda }\frac{\langle P|\overline\psi(z \hat z/2 +\vec b_\perp)\widetilde\Gamma\,{\cal W}_z\psi(-z \hat z/2)|\Omega\rangle}{\sqrt{Z_E(2L,b_\perp,Y=0)}}\nonumber
\end{align}
where $\lambda=zP^z$ and $\hat z^{\mu}=(0,1,\vec{0}_\perp)$;
$\widetilde\Gamma$ can be chosen as $\gamma_5\gamma^t$ or $\gamma_5\gamma^z$;
${\cal W}_z=W_z^{\dagger}(z \hat z/2 +\vec b_\perp)W_\perp W_z(-z \hat z/2)$ is a staple shaped gauge link along $-\hat z$ direction with $W_z(\xi)={\cal P}{\rm exp}\left[ig\int^{-L-\xi^z}_{0} ds A^z(\xi+s\hat z )\right]$ pointing to $-z$ direction.
Similar to Eq.~(\ref{eq:F_factorization}) and quasi-TMDPDF factorization~\cite{Ji:2019ewn}, $\widetilde \phi$ can be factorized into
a perturbative hard kernel and nonperturbative lightcone quantities~\cite{future}
\begin{align}\label{eq:phi_factorization}
\widetilde \phi(x,b_\perp,P)=H_\phi(x,P)\frac{\phi(x,b_\perp,P,\overline \delta)}{S_2(b_\perp,\delta,\overline \delta)}S_1(b_\perp,\delta,Y'=0)
\end{align}
where $\delta$ and $\overline \delta$ are on-lightcone regulators for gauge links along lightlike $P$ and its conjugate direction, and similar to Eq.~(\ref{eq:F_factorization}) $\delta$ and $\overline \delta$ are taking the lightcone limit implicitly.
The subscript ``1'' of the soft function denotes that one of the rapidity regulators are on-lightcone, and the other staple-shaped gauge link is along temporal direction indicated by $Y'=0$.
The soft functions $S_2$ and $S_1$
subtract away the regulator dependencies introduced in the lightcone TMDWF $\phi$.
The overall combination in the right hand side of Eq.~(\ref{eq:phi_factorization}) is rapidity
regularization scheme independent.

Combining Eqs.~(\ref{eq:F_factorization}) and (\ref{eq:phi_factorization}), we have
\begin{align}\label{eq:SSS}
&\frac{F(b_\perp,P\!\cdot\!P')}{\int dx dx' H(x,x',P,P')\widetilde\phi(x',b_\perp,P')\widetilde\phi^\dagger(x,b_\perp,P)}\nonumber \\
&=\frac{S_2(b_\perp,\delta,\delta')}{S_1(b_\perp,\delta,Y'=0) S_1(b_\perp,Y=0,\delta')}\equiv S_I(b_\perp)
\end{align}
where $H\equiv H_F(x,x')/H_\phi(x)H_\phi(x')$ is entirely
perturbative, and $S_I$ is called the intrinsic soft function. Similar to argument in Ref.~\cite{Collins:2011zzd}, the lightcone singularities cancel in the above combination, therefore the
result $S_I$ is scheme independent.
It is worth to point out that $S_I=e^{-\cal D}$ in the off-lightcone scheme from Eq.~(\ref{eq:S_diff_scheme}) even though $\cal D$ is scheme dependent in general.
The soft functions with on-lightcone regulator have the asymptotic forms for small $\delta$ similar to Eq.~(\ref{eq:S_diff_scheme})
\begin{align}\label{eq:S_on-lightcone}
S_1(b_\perp,\delta,Y')&=e^{(Y'-\ln \delta)K(b_\perp)+{\cal D}_1(b_\perp)+{\cal O}(\delta \exp(-Y))}\\
S_2(b_\perp,\delta,\delta')&=e^{-(\ln\delta\delta')K(b_\perp)+{\cal D}_2(b_\perp)+{\cal O}(\delta\delta')}\, .
\end{align}
Based on Eq.~(\ref{eq:SSS}), the Collins-Soper kernels $K$ are cancelled on the left hand side and we obtain the relation $2{\cal D}_1-{\cal D}_2={\cal D}$.
We have explicitly verified this relation and Eq.~(\ref{eq:SSS}) at one-loop level.

Similar to Eq.~(\ref{eq:SSS}), we can show that the cross section of DY can be factorized by quasi-TMDPDF~\cite{future}
\begin{align}\label{eq:sigma_DY}
\frac{d\sigma_{\rm DY}}{d^2 Q_\perp}&=\int dx\, dx' d^2 b_\perp e^{i\vec b_\perp \cdot \vec Q_\perp}\\
&\times\hat \sigma(x,x',P\!\cdot\! P')\widetilde f(x,b_\perp,P)\widetilde f(x',b_\perp,P')S_I(b_\perp)\nonumber
\end{align}
where $Q_\perp$ is the transverse momentum of produced lepton pair, $\hat \sigma(x,x',P\!\cdot\! P')$ is the hard kernel, and
\begin{align}
&\widetilde f(x,b_\perp,P)\\
&=\lim_{L\to\infty}\int\frac{d\lambda}{4\pi}e^{ix \lambda}\frac{\langle P|\overline\psi(z \hat z/2+\vec b_\perp)\widetilde \Gamma\,{\cal W}_z\psi(-z \hat z/2)|P\rangle}{P^z\sqrt{Z_E(2L,b_\perp,Y=0)}}\nonumber
\end{align}
is a quasi-TMD parton distribution with definition similar to the quasi-TMDWF in Eq.~(\ref{eq:quasi-TMDWF}).
With Eq.~(\ref{eq:sigma_DY}), the DY process in low-transverse-momentum region becomes predictable from first-principle calculations.\\

{\it Discussion and conclusion.}---To implement an actual calculation
of the soft function on lattice, particularly in the HQET framework,
some special considerations need to be made~\cite{Aglietti:1993hf,Hashimoto:1995in,Horgan:2009ti}.
It is known that the na{\"i}ve infinite heavy quark mass limit causes doubling problem,
and the usual technique, such as a Wilson term, can be used to lift the degeneracy.
The UV divergences from the transition current require renormalization, which
can be matched to the dimensional-regularization scheme, and the velocity
also need to be renormalized due to lattice artifacts. Moreover,
working with large velocity color sources might have similar challenges
as large-momentum hadrons~\cite{Bali:2016lva}.
For the soft function from the light-meson form factor, various
renormalization and matching will also need be made.
For correlators containing staple-shaped gauge links, the nonperturbative renormalization has been discussed in Refs.~\cite{Ji:2019ewn,Shanahan:2019zcq,Ebert:2019tvc}.
We reserve a detailed discussion about practicality of lattice calculations
to \cite{future}.

It shall be remarked that the soft function in the off-lightcone scheme approaches the lightcone
limit through the large rapidity separation $\sqrt{(2v\cdot v')^2/v^2(v')^2}=\sqrt{v^+v'^-/(v^-v’^+)}\to \infty$
but not through $v^2, (v')^2 \rightarrow 0$.

A common definition of the universal soft function was proposed in Refs.~\cite{Collins:2011ca,Collins:2011zzd}. The spacelike vectors $u^\mu=\gamma(\beta,1,0,0)$ and $u'^\mu=\gamma'(-\beta',1,0,0)$ were chosen instead of timelike $v $ and $v'$ to define the soft function for the DY process.
Despite the different definitions, we can prove that this soft function is equal to what we defined in Eq.~(\ref{eq:S_HQ_mu})~\cite{future}.

There are other efforts to propose soft functions on lattice connecting quasi-TMDPDF to lightcone TMDPDF~\cite{Ji:2014hxa,Ji:2018hvs,Ebert:2019okf}.
However, the soft function is controlled by cusp anomalous dimension at large hyperbolic angle, while other proposed soft functions are composed by Euclidean gauge links with circular angle which cannot be arbitrarily large.
Although, the cusp anomalous dimension of the bent soft functions in Ref.~\cite{Ji:2014hxa,Ebert:2019okf} coincide with $S_I$ at one-loop level, in general it is different beyond one-loop order~\cite{Korchemsky:1987wg}.
The difference in the cusp anomalous dimension will lead to different logarithmic structure $\ln\mu^2b_\perp^2$ which is not properly controlled by perturbation theory at large $b_\perp$.
As we have emphasized, $S_I$ can not be represented as a Wilson-loop composed of a finite number of Euclidean Wilson lines.

In conclusion, we find a Euclidean formulation of TMD soft function, which is a cross-section of color-charges moving along two conjugate lightlike directions that captures soft gluon effects in high energy process in QCD.
The soft function in off-lightcone regularization allows an interpretation as a form factor of a fast-moving heavy-quark pair in HQET.
The form factor and quasi-TMDWF of light meson can also be used to extract the intrinsic soft function $S_I$.
The results opens an opportunity to calculate soft functions nonperturbatively from lattice QCD and other Euclidean methods.
TMD factorization using quasi-TMD parton distribution together with intrinsic soft function $S_I$ allows the cross section of the DY process in low-transverse-momentum region to be predicted by first-principle calculations.

{\it Acknowledgment.}---We thank Stefan Meinel, Andreas Sch\"afer, Peng Sun, Wei Wang, Yi-Bo Yang, Feng Yuan, and Yong Zhao for valuable discussions.
This work is supported partially by Science and Technology Commission of Shanghai Municipality (Grant No.16DZ2260200), National Natural Science Foundation of China (Grant No.11655002 and No.11905126), and the US DOE grant DE-FG02-93ER-40762.

\bibliographystyle{apsrev4-1}
\bibliography{bibliography}

\end{document}